\documentclass{iau} 
\usepackage{graphicx}

\def\kms{\,km\,s$^{-1}$}

\def\hb{{\sc{H}}$\beta$\/}

\def\feii{Fe{\sc{ii}}}
\def\rfe{R$_{\rm{FeII}}$}

\def\mbh{$M\mathrm{_{BH}}$}

\def\LLEdd{$L\mathrm{_{bol}}/L\mathrm{_{Edd}}$}

\def\msun{M$_{\odot}$}

\def\rblr{$R\mathrm{_{BLR}}$}

\title[\feii{} strength in NLS1s] 
{\feii{} strength in NLS1s -- dependence on the viewing angle and FWHM }

\author[Swayamtrupta Panda, Paola Marziani \& Bo\.zena Czerny]   
{Swayamtrupta Panda$^{1,2}$, Paola Marziani$^3$ \and Bo\.zena Czerny$^1$}

\affiliation{$^1$Center For Theoretical Physics, Polish Academy of Sciences, Al. Lotnik\'ow 32/46, 02-668 Warsaw, Poland \\ email: {\tt panda@cft.edu.pl} \\[\affilskip]
$^2$Nicolaus Copernicus Astronomical Center, Polish Academy of Sciences, ul. Bartycka 18, 00-716 Warsaw, Poland \\
$^3$ INAF-Astronomical Observatory of Padova, Vicolo dell'Osservatorio, 5, 35122 Padova PD, Italy}

\pubyear{2019}
\volume{356}  
\jname{IAU356: Nuclear Activity in Galaxies across Cosmic Time}
\editors{M. Povic et al., eds.}
\begin{document}

\maketitle

\begin{abstract}

We address the effect of orientation of the accretion disk plane and the geometry of the broad line region (BLR) in the context of understanding the distribution of quasars along their Main Sequence. We utilize the photoionization code CLOUDY to model the BLR, incorporating the `un-constant' virial factor. We show the preliminary results of the analysis to highlight the co-dependence of the Eigenvector 1 parameter, \rfe{} on the broad \hb{} FWHM (i.e. the line dispersion) and the inclination angle ($\theta$), assuming fixed values for the Eddington ratio (\LLEdd{}), black hole mass (\mbh{}), spectral energy distribution (SED) shape, cloud density (n$\rm{_{H}}$) and composition.\footnote{The project was partially supported by NCN grant no. 2017/26/\-A/ST9/\-00756 (MAESTRO  9) and MNiSW grant DIR/WK/2018/12. PM acknowledges the INAF PRIN-\-SKA 2017 program 1.05.01.88.\-04.}

\end{abstract}

\firstsection 
\section{Introduction}
Narrow-Line Seyfert 1 (NLS1) galaxies constitute a typical class of Type-1 active galaxies which have ``narrow'' broad profiles (e.g. FWHM(\hb{}) $<$ 2000 \kms{}) and contain supermassive black holes (BH) that have masses lower than the typical broad-line Seyfert galaxies. Having a lower-than typical mass is a result of how the black hole mass (\mbh{}) is estimated. One of the methods to estimate the \mbh{} is the dynamical method that involves the use of the virial relation. According to this, the \mbh{} is a function of (i) the size of the emitting region, here, the broad-line region (BLR); and (ii) the FWHM of the virialized gas. The size of the BLR (i.e. \rblr{}) is measured by estimating the light-travel time from the central ionizing source to the emitting medium. This method is known as reverberation mapping (\cite[Peterson 1993]{peterson93}). The other quantity, the line FWHM, can be measured reliably from high quality spectroscopy. 

Since, the emitting line regions of quasars are extended, the energy that an observer receives from these luminous objects is also dependent on the geometry of the source wrt the observer. By geometry, we mean the structure and how this structure is oriented to the observer's line of sight. We address this aspect of the geometry of the quasars using photoionisation modelling with CLOUDY in the context of understanding better the main sequence of quasars (see \cite[Panda et al. 2019a]{Panda19a}, \cite[Panda et al. 2019b]{Panda19b} for more details).

The eigenvector 1 of the original principal component analysis (PCA) paved way for the quasar main sequence picture as we know it today (\cite[Sulentic et al. 2000]{sul00}, \cite[Shen \& Ho 2014]{sh14}). The main sequence connects the velocity profile of `broad' H$\beta$ with the strength of the \feii\ emission (\rfe), i.e., the intensity of the \feii\ blend within 4434-4684 \AA\ normalized with the `broad' H$\beta$ intensity.

\section{Method}

\begin{figure}
\begin{center}
 \includegraphics[width=\columnwidth]{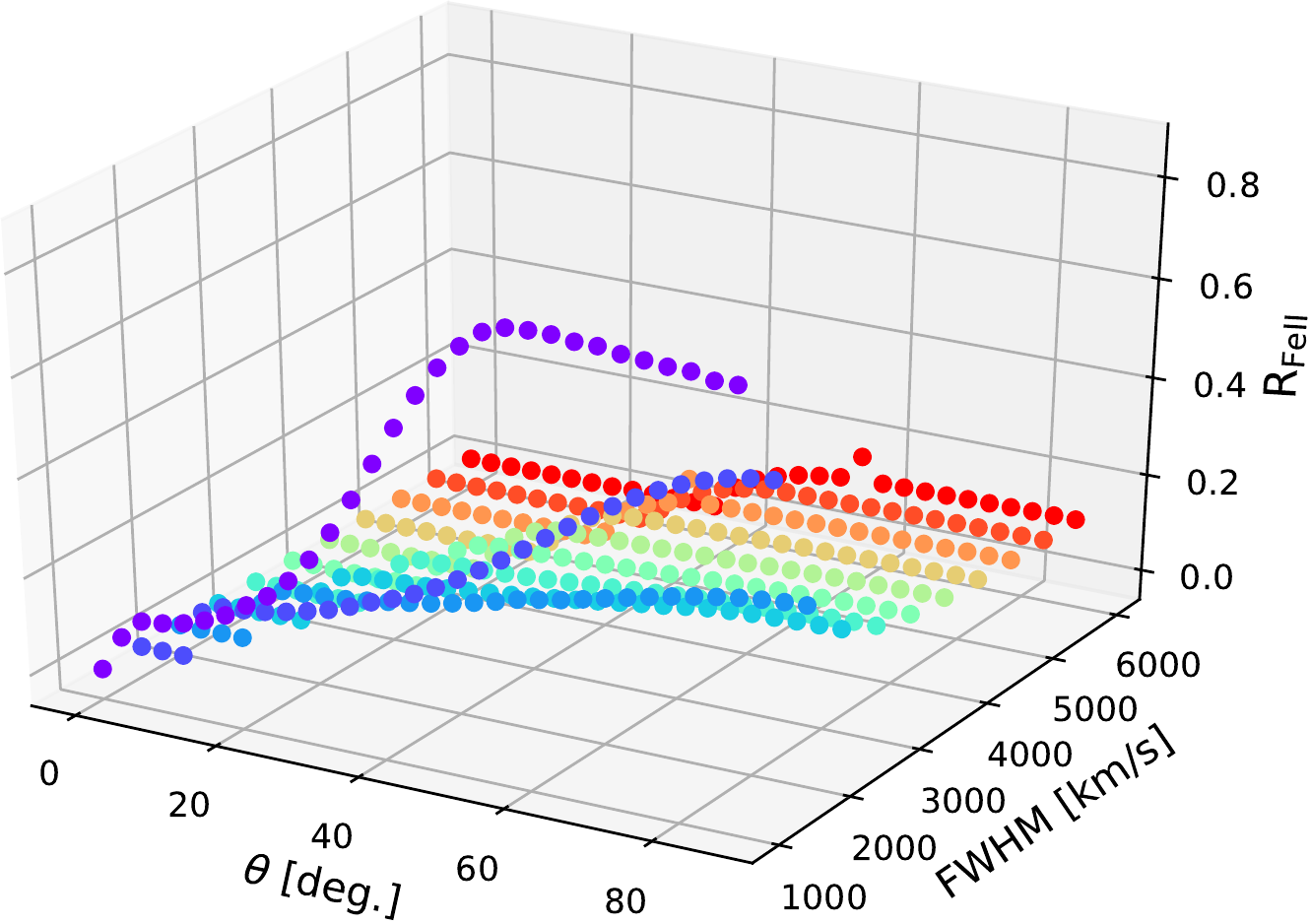} 
 \caption{3D scatter plot showing the dependence of the parameter \rfe{} on the \hb{} FWHM and the inclination angle ($\theta$). The \hb{} FWHM ranges from 1000 \kms{} to 6000 \kms{} with a step size of 500 \kms{}. Similarly, the $\theta$ values range from 0$^{\rm{o}}$-90$^{\rm{o}}$ with a step size of 3$^{\rm{o}}$. The black hole mass is assumed to be 10$^8$\msun{} and the Eddington ratio, \LLEdd{} = 0.25. The value of the $\kappa$ = 0.1 consistent with a flat, keplerian-like gas distribution around the central supermassive black hole.} 
   \label{fig1}
\end{center}
\end{figure}

We assume a single cloud model where the density ($n_\mathrm{H}$) of the ionized gas cloud is varied from $10^{9}\; \mathrm{cm^{-3}}$ to $10^{13}\; \mathrm{cm^{-3}}$ with a step-size of 0.25 (in log-scale). We utilize the \textit{GASS10} model \cite[Grevesse et al. (2010)]{gass10} to recover the solar-like abundances and vary the metallicity within the gas cloud, going from a sub-solar type (0.1 Z$_{\odot}$) to super-solar (100 Z$_{\odot}$) with a step-size of 0.25 (in log-scale). The total luminosity of the ionizing continuum is derived assuming a value of the Eddington ratio (\LLEdd{}) and the respective value for the black hole mass (here, we assume an \LLEdd{} = 0.25 and a \mbh{} = 10$^8$ \msun{}). The Eddington ratio is appropriate for sources of Population A. The \mbh{} is representative of optically-selected low-z quasar samples. The shape of the ionizing continuum used here is taken from \cite[Korista et al. (1997)]{kor97}. The size of the BLR is estimated from the virial relation, assuming a black hole mass, a distribution in the viewing angle [0-90 degrees] and FWHM which is given as

\begin{equation}
    R_{BLR} = \frac{GM_{BH}}{f*FWHM^2}
\end{equation}

where G is the Gravitational constant. The \textit{f} factor, which contains the information about the geometry of the source, can be expressed as  

\begin{equation}
    f = \frac{1}{4\left[\kappa^2 + sin^2\theta\right]}
\end{equation}

where, $\theta$ is the angle of inclination wrt the observer and $\kappa$ is the ratio between $v_\mathrm{iso}$ and $v_\mathrm{K}$, which decides how isotropic the gas distribution is around the central potential. If the value is close to zero, it represents a flat disk with thickness almost zero. On the other hand, if the value of $\kappa$ is close to unity, it represents an almost spherical distribution of the gas. Here, we assume $\kappa$ = 0.1 that is consistent with a flat, keplerian-like gas distribution.

Substituting the values for the \mbh{} and $\kappa$, we have

\begin{equation}
    R_{BLR} \approx 5.31\times 10^{24}\left[\frac{0.01 + sin^2\theta}{FWHM^2}\right] \quad\quad\quad(\rm{in\;cm})
    \label{eq:3}
\end{equation}

\section{Results and Conclusions}
Figure \ref{fig1} shows dependence of the \rfe{} both on FWHM and $\theta$. \rfe{} values are anti-correlated with the FWHM. This is directly coming from the Eq. \ref{eq:3}, where the \rblr{} is anti-correlated with the square of the FWHM, but also correlates with $\theta$ which shows that if \rblr{} is too low, the ionizing radiation is too high for the \feii{} species to exist and the emitting zones of the \feii{} only begins to materialize deeper in the cloud (see \cite[Panda et al. 2018]{Panda18}). This effect reduces the \rfe{} for lower \rblr{} values. $\theta$ is $<$ 60$^{\rm{o}}$ for Type-1 quasars. Within this limit, we find an increasing trend in \rfe{} wrt $\theta$. This trend is more substantial for lower values of the FWHM that are consistent with NLS1 population, i.e. $\lesssim$ 3000 \kms{}. These results are described and analyzed in \cite[Panda et al. (2019c)]{Panda19c}.

\end{document}